\documentclass[sigplan,nonacm]{acmart}
\AtBeginDocument{%
  }

\setcopyright{acmlicensed}
\copyrightyear{2018}
\acmYear{2018}
\acmDOI{XXXXXXX.XXXXXXX}
\acmConference[Conference acronym 'XX]{Make sure to enter the correct
  conference title from your rights confirmation email}{June 03--05,
  2018}{Woodstock, NY}
\acmISBN{978-1-4503-XXXX-X/2018/06}

\usepackage{amsmath}
\usepackage{graphicx}
\usepackage{textcomp}
\usepackage{xcolor}
\usepackage{hyperref}
\usepackage{array}
\usepackage{booktabs} 
\usepackage{array}
\usepackage{tabularx}
\usepackage[utf8]{inputenc}
\usepackage{graphicx}
\usepackage{subcaption}
\usepackage{multirow}
\usepackage{amsmath}
\usepackage{tcolorbox}
\usepackage{tcolorbox}
\usepackage{tikz}
\usetikzlibrary{positioning, calc}
\usepackage[table]{xcolor}

 \usepackage{amsmath}
  \usepackage{algorithm}       
  \usepackage{algpseudocode}   
  \algrenewcommand{\algorithmiccomment}[1]{\hfill\footnotesize$\triangleright$~#1}

\begin{document}

\title[]{PRISM: Probabilistic Runtime Insights and \\ Scalable Performance Modeling for Large-Scale Distributed Training}


\author{
Alicia Golden$^{\dagger\ddagger}$, Michael Kuchnik$^\dagger$, Samuel Hsia$^\dagger$, Zachary DeVito$^\dagger$,\\
Gu-Yeon Wei$^\ddagger$, David Brooks$^\ddagger$, Carole-Jean Wu$^\dagger$\\
$^\dagger$FAIR at Meta, $^\ddagger$Harvard University
}

\renewcommand{\shortauthors}{}

\begin{abstract}

Large model training beyond tens of thousands of GPUs is an uncharted territory. At such scales, disruptions to the training process are not a matter of if, but a matter of when—a stochastic process degrading training productivity. Dynamic runtime variation will become increasingly more frequent as training scales and GPUs are operated in increasingly power-limited and thermally-stressed environments. At the 64,000+ GPU scale, we already observe 12\% variability for frontier foundation model training. Motivated by our analysis and the large design space around performance variability, we present PRISM — a performance modeling framework that captures the stochastic nature of large-scale distributed training. The core of PRISM is a statistical model that composes operator-level latency distributions through workload dependencies. Across 14 diverse training configurations spanning hundreds to 64K+ GPUs, PRISM estimates p95 execution time within 5.4\% error. Using PRISM, we explore the design and optimization space of distributed training, enabling principled, variability-aware recommendations that can improve performance and system efficiency at scale.
\end{abstract}


\maketitle

\section{Introduction}

Distributed training is outgrowing the assumption that performance is predictable. Training modern foundation models requires the coordination of thousands of GPUs across complex datacenter environments~\cite{meta2025llama4}, subject to a variety of external stresses. Runtime variation can begin at the level of the individual device, as even seemingly identical hardware exhibits variability. On the compute side, clock frequency and DVFS behavior~\cite{clockfreq}, thermal
throttling~\cite{gatech}, and silicon process variation introduced by GPU binning~\cite{binning} cause otherwise identical devices to diverge. For communication,
NVSwitch contention~\cite{nvlink_congestion} can perturb intra-node transfers, while network congestion~\cite{network_congestion} can disrupt inter-node ones. Importantly, these effects are not due to failure, but can occur during otherwise healthy runs. 

\begin{figure}[t]
    \centering    \includegraphics[width=\linewidth]{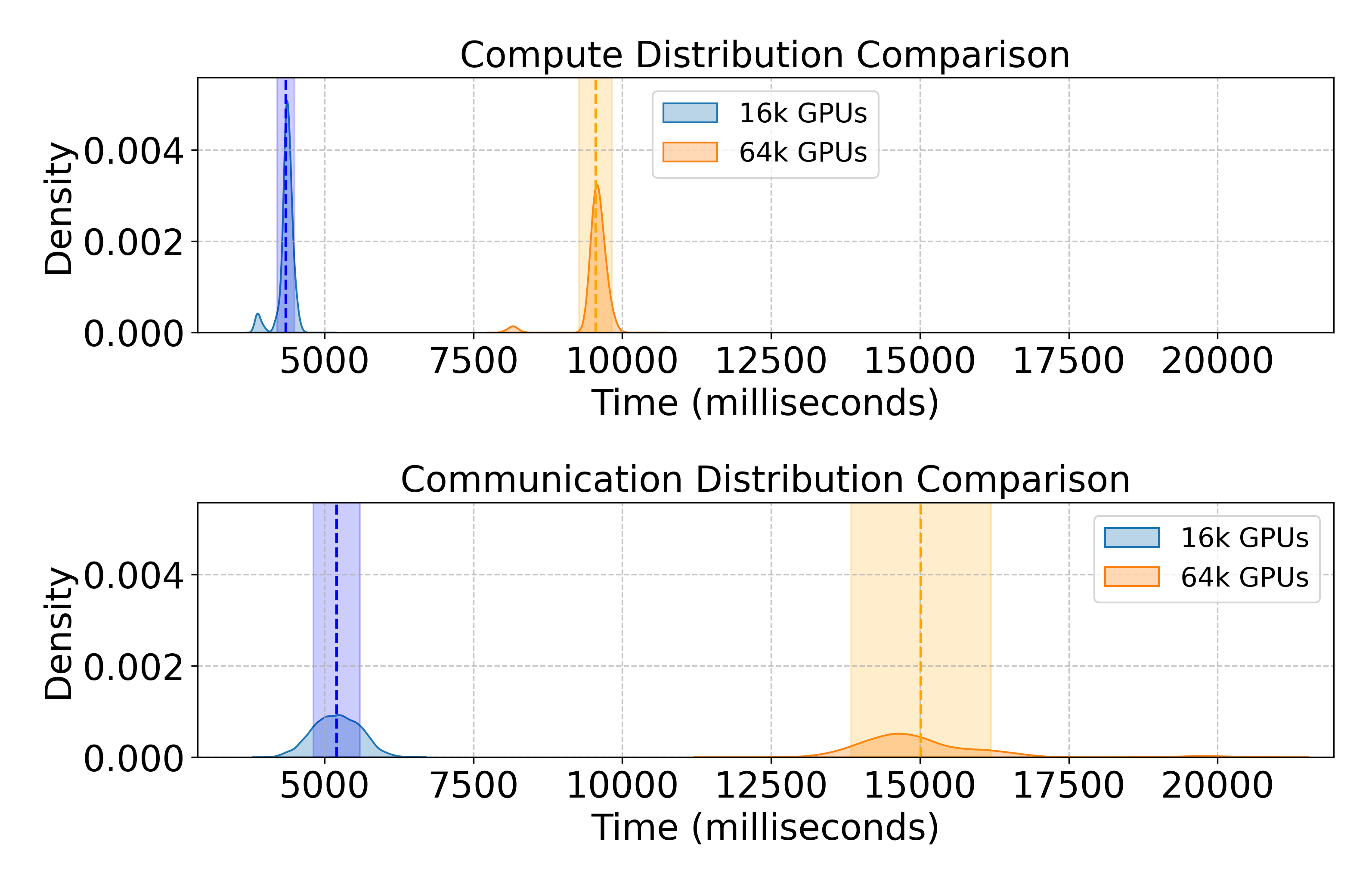}
    \caption{Distribution of compute and communication time for two LLM training experiments in real-world production training environment -- across 16K and 64K+ NVIDIA GPUs respectively. We find the 64K+ job exhibits 2.03$\times$ higher standard deviation in compute and 2.91$\times$ higher standard deviation in communication compared to the 16K job.}
    \label{fig:motivation}
\end{figure}

Runtime variability is especially costly because distributed training is synchronous. Since a training step completes only after all participating GPU ranks have finished, a slowdown in one rank can stall all others that finish faster. These GPUs remain idle for the remainder of the step, directly reducing utilization. As a result, step time is governed not by typical execution behavior, but rather by the slowest rank, often called a straggler~\cite{litsilicon}.

Scale compounds the problem. Each step in a larger run contains more operations and participating ranks, creating higher probability of a slow tail. We observe this effect directly in production training. Figure~\ref{fig:motivation} compares compute and communication times from two LLM training jobs running on 16K and 64K+ GPUs. The 64K+ job exhibits 2.03$\times$ higher standard deviation in compute time and 2.91$\times$ higher standard deviation in communication time per step. Additionally across training steps, we observe 12\% variation between p95/p50 for the 64K+ GPU run. These results show that runtime variability is not an isolated anomaly at frontier scale, but rather a reoccurring property of distributed execution.

Despite this, existing performance models generally assume deterministic execution~\cite{zhang2024megascale,metis,amped}. They assign a fixed duration to each operation and produce a single estimate of training time. However, this abstraction cannot represent a runtime distribution, determine how frequently slow ranks delay synchronization, or capture how operation-level variation propagates through the critical path. Since a training step ends only when the slowest rank finishes, deterministic models systematically miss the tail behavior that actually governs step time.

At the same time, empirically measuring this variability is often impractical at scale. Accurately characterizing a runtime distribution requires repeatedly executing the same training workload. Each run can occupy hundreds or thousands of GPUs, and design-space exploration would require repeating this process for every candidate hardware state, placement, or optimization. The measurement cost grows with both the scale of the workload and the number of configurations being considered. As a result, practitioners currently lack a practical way to estimate how fine-grained variability affects end-to-end performance or to determine which system decisions can mitigate its impact.

To address this gap, we introduce \textbf{PRISM}: \textbf{P}robabilistic \textbf{R}untime \textbf{I}nsights and \textbf{S}calable Performance \textbf{M}odeling for large-scale distributed training. PRISM is a framework that models training as a \textit{stochastic} process rather than a \textit{deterministic} one, and provides an efficient way to model variation at-scale. PRISM captures fine-grained variability at the level of individual compute and communication operations, represents execution as a dependency graph, and uses Monte Carlo simulation to estimate execution time distributions. 

We use p95 training-step time as PRISM’s primary modeling target, as tail performance is a direct concern for distributed training schedulers. Placement and resource-allocation decisions must account for the slow steps that reduce throughput and leave resources idle. While p50 captures typical execution, it hides slowdowns and is insensitive to behavior in the upper tail. At the other extreme, the maximum is highly sensitive to isolated events such as checkpointing or transient system stalls. p95 provides a useful middle ground: it captures recurring tail slowdowns that are relevant to scheduling decisions while remaining representative of the compute and communication variability that PRISM models. Across 14 diverse production runs, PRISM estimates p95 training step time within 5.4\% error. This demonstrates that PRISM's probabilistic modeling can accurately capture tail performance in distributed training systems.

PRISM additionally enables variability-aware system design by identifying how and where runtime variability matters most. To this end, we build PRISM-Guide, which evaluates candidate system configurations using PRISM and ranks them according to their tail-performance impact. In a production-inspired degraded-GPU placement study grounded in an H100 fleet of 286,212 GPUs, PRISM-Guide selects the measured-best pipeline stage for a GPU operating at 1200 MHz, avoiding up to an 11.47 percentage-point p95 penalty relative to worst placement. Extending this analysis further across five workloads and five degraded-GPU operating points, PRISM-Guide reduces the p95 penalty by up to 4.9--30.2 percentage points. PRISM-Guide also identifies which sources of variability are worthwhile optimization targets, since high kernel-level variance alone does not imply high system-level impact. Together, these studies turn PRISM's p95 estimates into actionable guidance for placement and performance optimization.

Overall, our work introduces a new approach to performance modeling for large-scale training,  treating runtime variability as a first-class property of the system. By estimating tail performance and tracing its causes through distributed execution, PRISM provides both an accurate estimate of runtime under real-world variability and a mechanism for identifying the design choices that most strongly affect it. This enables more predictable training and provides a foundation for variability-aware scheduling, placement, and performance optimization at scale. The main contributions of this work are as follows:

\begin{itemize}
    \item \textbf{We quantify variation across more than 
    20,000 GPUs across a real-world production datacenter fleet.} We find GEMM kernels can experience as much as 14\% variation, depending on training hardware and deployment environment.
    
    \item \textbf{We formulate end-to-end training performance estimation as a statistical problem}, motivate p95 as a practical tail-performance target, and capture how fine-grained variability in individual kernels and operations composes through execution dependencies into execution time.

    \item \textbf{We introduce PRISM --- a \textit{variability-aware} performance modeling framework for large-scale distributed training.}     PRISM moves beyond deterministic performance estimation to estimate p95 execution time under realistic hardware and system variability. Across 14 configurations spanning hundreds to 64K+ GPUs, PRISM estimates p95 training step time within 5.4\% error.

    \item \textbf{We leverage PRISM to enable variability-aware system design using PRISM-Guide,} which turns PRISM’s tail-performance estimates into design recommendations. In a measured degraded-GPU placement experiment, PRISM-Guide selects correct pipeline placement and avoids up to an 11.47 percentage point p95 penalty. Across an extended sweep of workloads and degradations, PRISM-Guide identifies placements that reduce p95 penalty by 4.9--30.2 percentage points.

\end{itemize}

The rest of the paper is organized as follows: Section 2 quantifies runtime variability in production clusters, Section 3 introduces the PRISM performance modeling framework, Section 4 outlines experimental methodology, Section 5 dives into PRISM's evaluation results across our 14 workload suite, and Section 6 leverages PRISM-Guide to explore variability-aware system design. Upon acceptance, we will open-source the PRISM framework to encourage future work in variability-aware optimization.

\section{Background and Motivation: Variability At-Scale}

\subsection{Sources of Variability in Real-World Training}

Training performance at-scale is highly dynamic, subject to variation factors that take many forms. While it is well-studied that failures during training can lead to excessive churn at the cluster level~\cite{kokolis2025reliability}, \textit{even failure-free runs are subject to performance variation}. There are two main types of variation observed. First, when the same computations are executed on the same hardware system repeatedly, the execution time can be a stochastic quantity -- i.e., \textit{temporal timing variability}. Second, when the same computations are run concurrently over a large collection of GPUs, hardware process variation can lead to further performance variability -- i.e., \textit{spatial timing variability}. Both compute and communication operations can exhibit temporal and spatial variability.

\definecolor{rootfill}{HTML}{EAEAEA}
\definecolor{rootstroke}{HTML}{333333}
\definecolor{compfill}{HTML}{E0CDD2}
\definecolor{compstroke}{HTML}{7A5A62}
\definecolor{comptext}{HTML}{4A3540}
\definecolor{compheader}{HTML}{ECDADF}
\definecolor{compbody}{HTML}{F8F2F4}
\definecolor{commfill}{HTML}{E8D5B7}
\definecolor{commstroke}{HTML}{B8860B}
\definecolor{commheader}{HTML}{F5E6CC}
\definecolor{commbody}{HTML}{FDF6EC}
\definecolor{itemcolor}{HTML}{333333}

\newcommand{\djtlf}{\fontfamily{DejaVuSans-TLF}\selectfont}
\newcommand{\taxrow}[2]{\makebox[6.3cm][s]{#1\hfill#2}}
\newcommand{\taxrownocite}[1]{\makebox[6.3cm][l]{#1}}

\begin{figure}[t]
\centering
\resizebox{\columnwidth}{!}{%
\begin{tikzpicture}[
    font=\djtlf,
    rownode/.style={
        anchor=west, font=\djtlf\large, text=itemcolor, inner sep=0pt,
    },
]

\fill[rootfill, draw=rootstroke, line width=1pt, rounded corners=3pt]
    (3.3, 0) rectangle (11.3, -1.0);
\node[font=\djtlf\bfseries\large, text=rootstroke]
    at (7.3, -0.5) {Latency Variation};

\draw[rootstroke!70, line width=1pt] (7.3,-1.0) -- (7.3,-1.3);
\draw[rootstroke!70, line width=1pt] (3.5,-1.3) -- (11.1,-1.3);
\draw[rootstroke!70, line width=1pt] (3.5,-1.3) -- (3.5,-1.55);
\draw[rootstroke!70, line width=1pt] (11.1,-1.3) -- (11.1,-1.55);

\fill[compfill, draw=compstroke, line width=1pt, rounded corners=3pt]
    (0, -1.55) rectangle (7.0, -2.35);
\node[font=\djtlf\bfseries\large, text=comptext]
    at (3.5, -1.95) {Compute};

\fill[commfill, draw=commstroke, line width=1pt, rounded corners=3pt]
    (7.6, -1.55) rectangle (14.6, -2.35);
\node[font=\djtlf\bfseries\large, text=comptext]
    at (11.1, -1.95) {Communication};

\draw[compstroke, line width=0.7pt] (3.5,-2.35) -- (3.5,-2.65);
\draw[commstroke, line width=0.7pt] (11.1,-2.35) -- (11.1,-2.65);


\fill[compbody, draw=compstroke, line width=0.7pt, rounded corners=3pt]
    (0, -2.65) rectangle (7.0, -4.75);
\fill[compheader, rounded corners=3pt]
    (0, -2.65) rectangle (7.0, -3.35);
\fill[compheader] (0, -3.27) rectangle (7.0, -3.35);
\draw[compstroke, line width=0.4pt] (0,-3.35) -- (7.0,-3.35);
\node[font=\djtlf\bfseries\large, text=comptext]
    at (3.5, -3.0) {Power \& Thermal};
\node[rownode] at (0.35, -3.75)
    {\taxrow{Clock frequency / DVFS}{\cite{clockfreq}}};
\node[rownode] at (0.35, -4.4)
    {\taxrow{Thermal throttling}{\cite{gatech}}};

\draw[compstroke, line width=0.7pt] (3.5,-4.75) -- (3.5,-5.05);

\fill[compbody, draw=compstroke, line width=0.7pt, rounded corners=3pt]
    (0, -5.05) rectangle (7.0, -6.45);
\fill[compheader, rounded corners=3pt]
    (0, -5.05) rectangle (7.0, -5.75);
\fill[compheader] (0, -5.67) rectangle (7.0, -5.75);
\draw[compstroke, line width=0.4pt] (0,-5.75) -- (7.0,-5.75);
\node[font=\djtlf\bfseries\large, text=comptext]
    at (3.5, -5.4) {Silicon Process Variation};
\node[rownode] at (0.35, -6.1)
    {\taxrow{GPU binning/yield variation}{\cite{binning}}};

\draw[compstroke, line width=0.7pt] (3.5,-6.45) -- (3.5,-6.75);

\fill[compbody, draw=compstroke, line width=0.7pt, rounded corners=3pt]
    (0, -6.75) rectangle (7.0, -10.25);
\fill[compheader, rounded corners=3pt]
    (0, -6.75) rectangle (7.0, -7.45);
\fill[compheader] (0, -7.37) rectangle (7.0, -7.45);
\draw[compstroke, line width=0.4pt] (0,-7.45) -- (7.0,-7.45);
\node[font=\djtlf\bfseries\large, text=comptext]
    at (3.5, -7.1) {Software / Runtime};
\node[rownode] at (0.35, -7.85)
    {\taxrow{CUDA scheduling}
    {\cite{threadCUDAscheduling}}};
\node[rownode] at (0.35, -9.15)
    {\taxrow{Cache accesses}
    {\cite{cacheaccesses}}};
\node[rownode] at (0.35, -8.5)
    {\taxrow{Garbage collection} 
    {\cite{lin2025stragglers}}};
\node[rownode] at (0.35, -9.8)
    {\taxrow{FP non-determinism}{\cite{golden2024flashattentionstable}}};


\fill[commbody, draw=commstroke, line width=0.7pt, rounded corners=3pt]
    (7.6, -2.65) rectangle (14.6, -5.45);
\fill[commheader, rounded corners=3pt]
    (7.6, -2.65) rectangle (14.6, -3.35);
\fill[commheader] (7.6, -3.27) rectangle (14.6, -3.35);
\draw[commstroke, line width=0.4pt] (7.6,-3.35) -- (14.6,-3.35);
\node[font=\djtlf\bfseries\large, text=comptext]
    at (11.1, -3.0) {Intra-Node};
\node[rownode] at (7.95, -3.75)
    {\taxrow{NVSwitch contention}{\cite{nvlink_congestion}}};
\node[rownode] at (7.95, -4.4)
    {\taxrow{NUMA topology effects}{\cite{chrapek2024exploring}}};
\node[rownode] at (7.95, -5.05)
    {\taxrow{NIC degradation}{\cite{wu2025greyhound}}};

\draw[commstroke, line width=0.7pt] (11.1,-5.45) -- (11.1,-5.75);

\fill[commbody, draw=commstroke, line width=0.7pt, rounded corners=3pt]
    (7.6, -5.75) rectangle (14.6, -10.0);
\fill[commheader, rounded corners=3pt]
    (7.6, -5.75) rectangle (14.6, -6.45);
\fill[commheader] (7.6, -6.37) rectangle (14.6, -6.45);
\draw[commstroke, line width=0.4pt] (7.6,-6.45) -- (14.6,-6.45);
\node[font=\djtlf\bfseries\large, text=comptext]
    at (11.1, -6.1) {Inter-Node};
\node[rownode] at (7.95, -6.85)
    {\taxrow{Network congestion}{\cite{network_congestion}}};
\node[rownode] at (7.95, -7.55)
    {\taxrow{IB / RoCE congestion}{\cite{subramanya2024cassini}}};
\node[rownode] at (7.95, -8.25)
    {\taxrow{ECMP hash imbalance}{\cite{zhai2024alibaba}}};
\node[rownode] at (7.95, -8.95)
    {\taxrow{Adaptive routing delays}{\cite{chrapek2024exploring}}};
\node[rownode] at (7.95, -9.65)
    {\taxrow{Switch buffer contention}{\cite{switchbuffer}}};

\end{tikzpicture}%
}
\caption{Taxonomy of potential sources of latency variation in distributed training \textit{(non-exhaustive)}. A wide range of compute and communication factors can contribute to aggregated variability in distributed training at-scale.}
\label{fig:taxonomy_variation}
\end{figure}

To better understand the sources of such variation, we construct a taxonomy to categorize the potential root causes of compute and communication variation. Figure~\ref{fig:taxonomy_variation} shows this outline, though we note it is a non-exhaustive list. On the compute side, clock frequency variation, thermal throttling, and silicon process variation can cause GPU performance to diverge even across nominally identical devices~\cite{clockfreq, gatech,xiong2024superbench}. Software-level factors such as CUDA scheduling and floating-point non-determinism~\cite{golden2024flashattentionstable, threadCUDAscheduling} can also introduce additional noise. On the communication side, intra-node variation can stem from NVSwitch contention and NIC degradation~\cite{chrapek2024exploring,wu2025greyhound}, while inter-node variation is driven by network congestion, ECMP hash imbalance, and buffer contention~\cite{zhai2024alibaba,subramanya2024cassini}. 

\subsection{Quantifying Variability in Production Clusters}

\begin{figure}[t]
    \centering
    \begin{subfigure}{\linewidth}
        \centering
        \includegraphics[width=\linewidth]{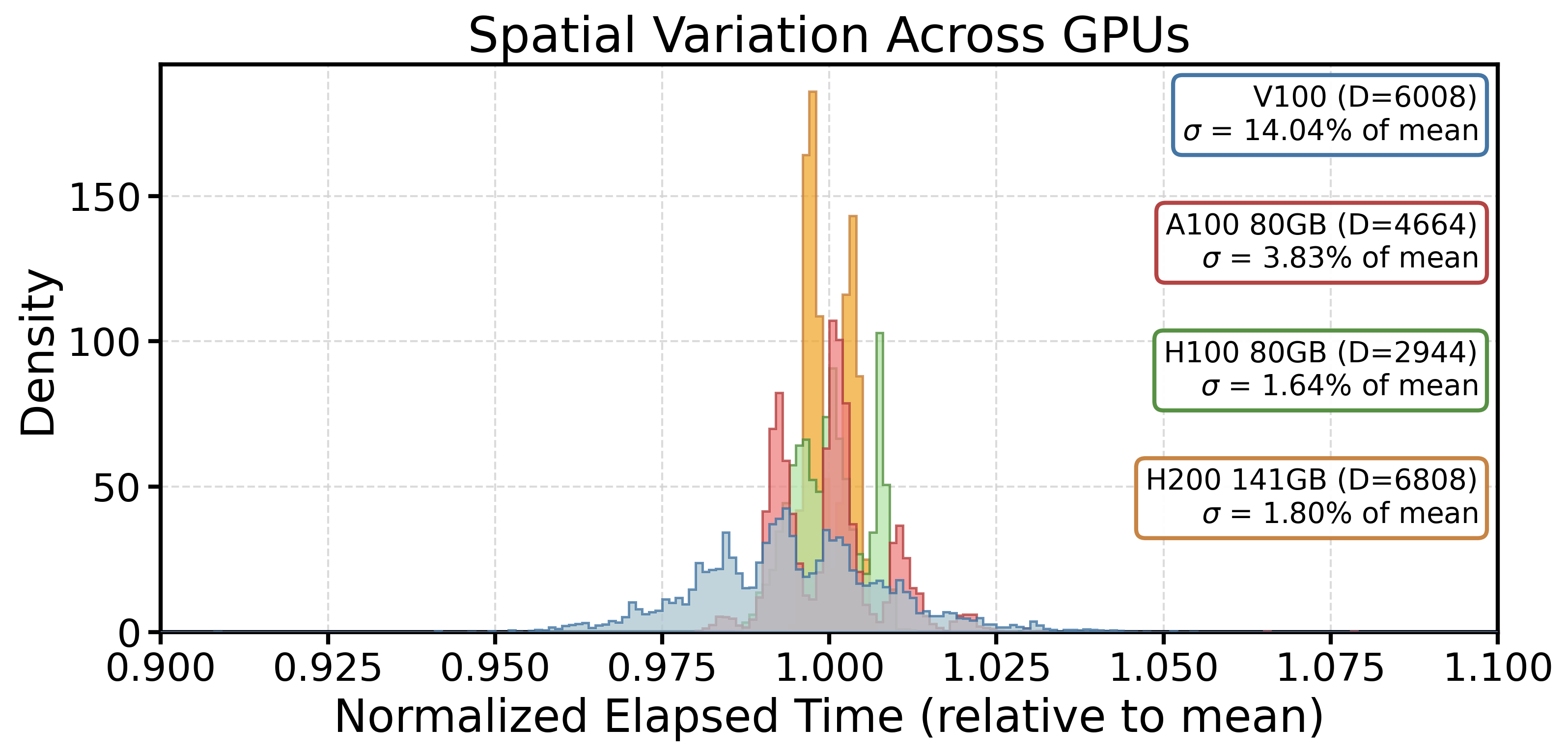}
        \caption{Spatial Timing Variability. Each GEMM kernel is run across D=2944-6808 unique GPUs across our clusters. We sample the p50 average value on each GPU and find there exists 1.64-14.04\% spatial variation.}
        \label{fig:var-subfigA}
    \end{subfigure}
    \medskip 
    \begin{subfigure}{\linewidth}
        \centering
         \includegraphics[width=\linewidth]{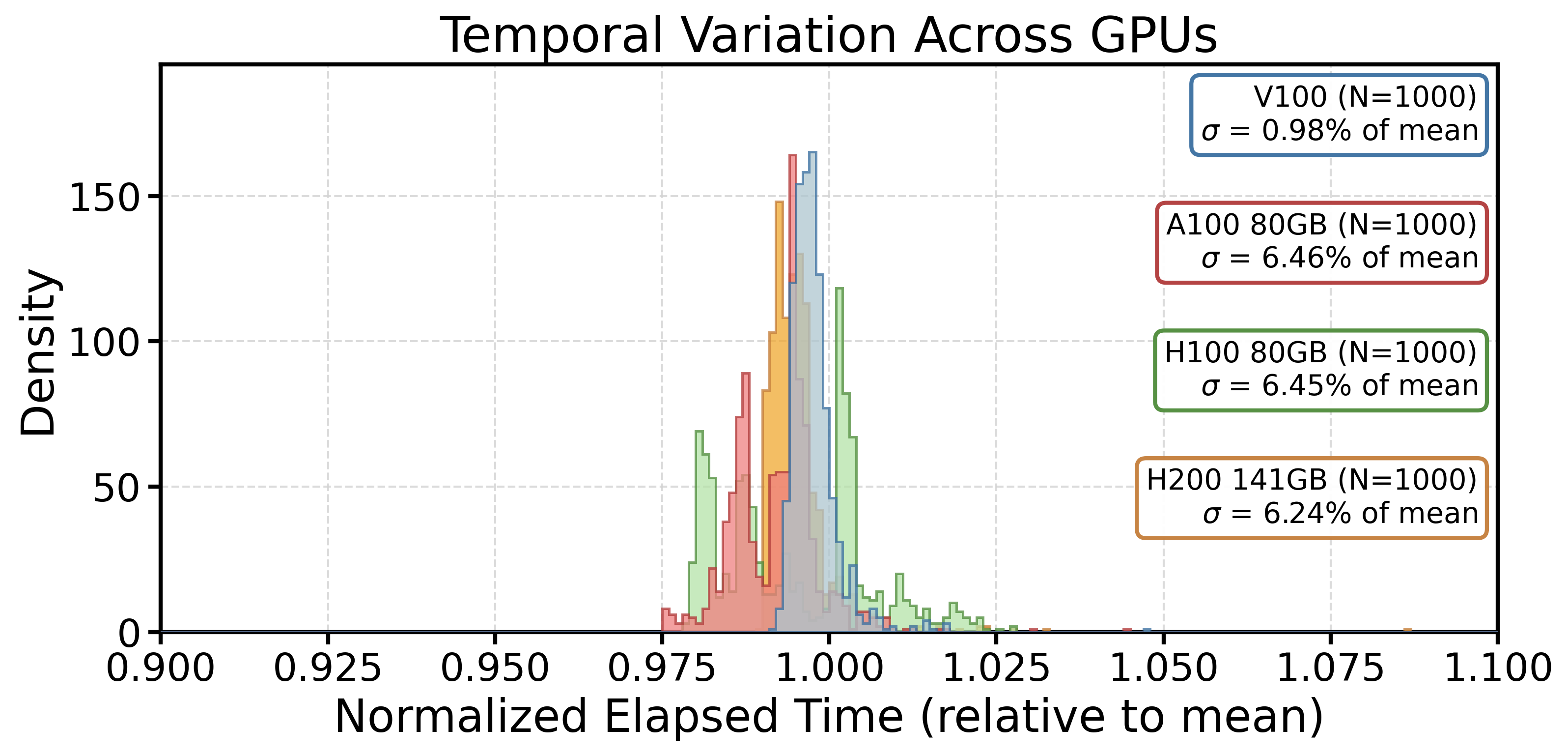} 
        \caption{Temporal Timing Variability. We profile N=1000 iterations of an identical GEMM kernel on the p50 representative GPU across our clusters and find 0.98-6.46\% variation.}
        \label{fig:var-subfigB}
    \end{subfigure}
    \caption{Significant latency variability is observed for GPUs over the fleet (spatial variability) and the same computation kernel on identical GPUs (temporal variability).}
    \label{fig:variation_kernel}
\end{figure}

\begin{figure}[!h]
    \centering
        \centering
         \includegraphics[width=\linewidth]{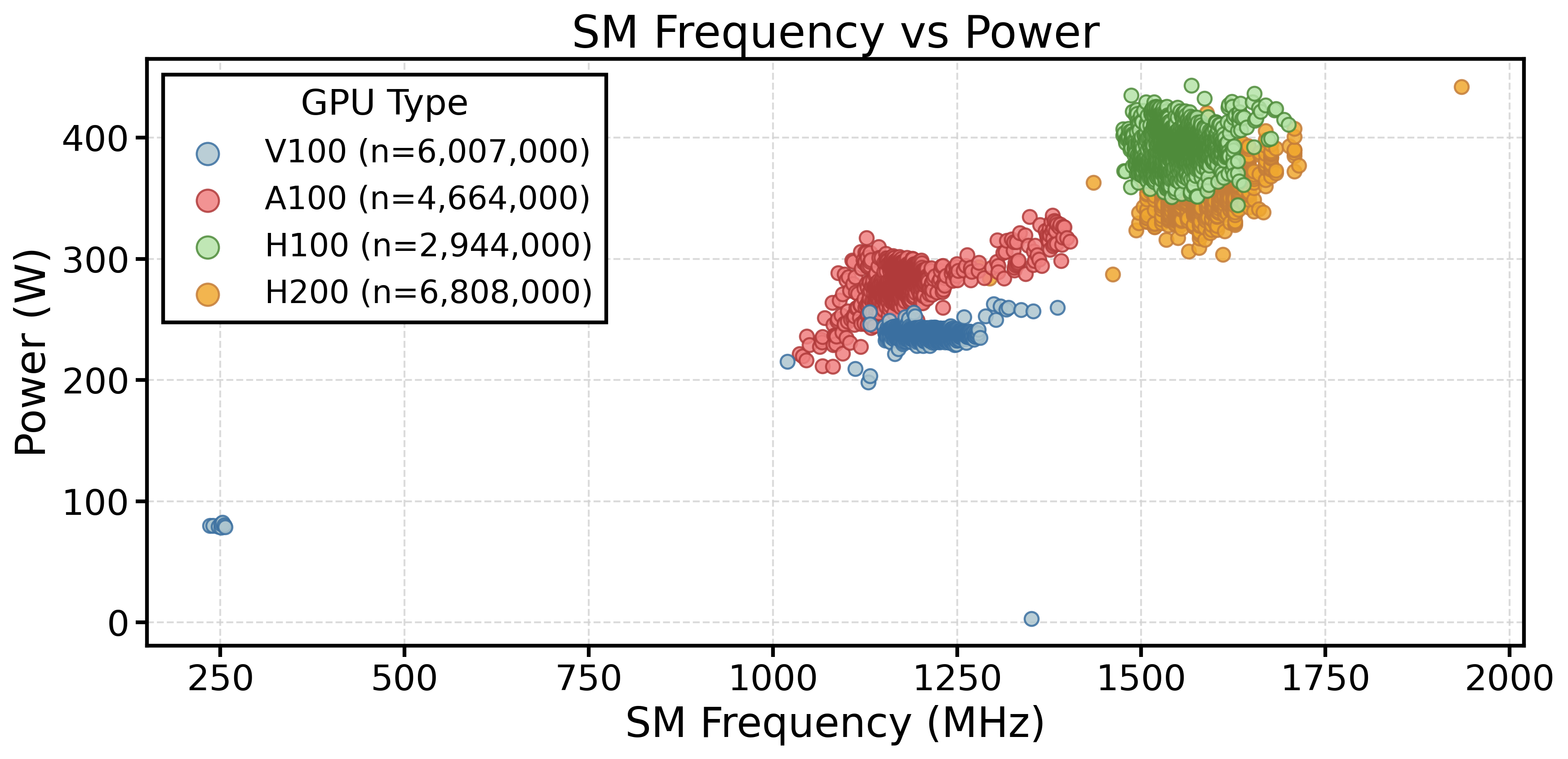} 
        \caption{One source of latency variability comes from the dynamically varying frequency setting of GPUs. We observe frequency and power distribution over GEMM kernels when measuring spatial variation. A100s see 500 MHz difference in frequency, corresponding to 140W power differential.
        }
        \label{fig:variation_sm_subfigB}
    \label{fig:variation_sm}
\end{figure}

\begin{figure*}[tb]
    \centering    \includegraphics[width=\linewidth]{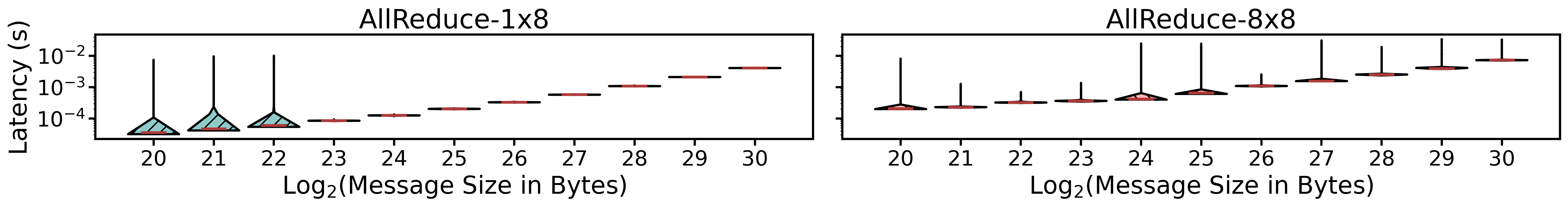}
    \caption{Communication primitive distributions. AllReduce profiled on both one and eight boxes of NVIDIA-H200 GPUs across intra-node NVLink \textit{(left)}, and inter-node backend network  
    \textit{(right)}, respectively. Intra-node communication is nearly deterministic at large message sizes, though we measure rare stragglers at small message sizes. In contrast, inter-node traffic is more variable, with latency fluctuating at millisecond level and tail latency spanning up to an order of magnitude.}
    \label{fig:comm_collectives}
\end{figure*}

\begin{figure}[t]
    \centering
    
        \includegraphics[width=1.0\linewidth]{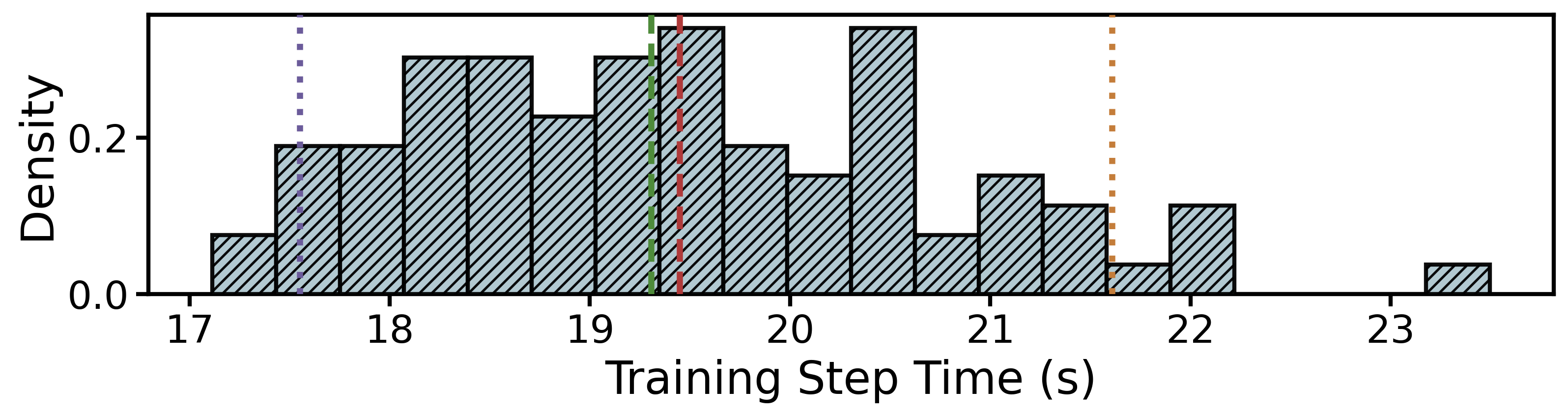} 
        \caption{Observed training step-time variability across production run. Distribution of training step times for the 64K+ GPU run. We find p5/p50/p95 of 17.6/19.3/21.6 seconds, a 12\% spread from median to p95. }
        \label{fig:variation_64k_subfigB}
    \label{fig:variation_64k}
\end{figure}

To assess the scale of this variation, we characterize three generations of real production training clusters, each equipped with tens of thousands of GPUs. We  design microbenchmarks to carefully characterize the \textit{temporal timing variability} and \textit{spatial timing variability} of real production training clusters across GPU generations -- including NVIDIA V100, A100, H100, and H200, spanning a release range across 7 years.

We begin by quantifying \textit{spatial timing variability} across more than 20,000 unique NVIDIA GPUs in our production clusters. Figure \ref{fig:var-subfigA} shows the observed distributions representing spatial variability across the fleet with time on the x-axis. We find there exists 1.64-14.04\% spatial variation across GPUs for the GEMM kernel.

We then quantify \textit{temporal timing variability} of the GPUs, using the p50 representative GPU across our clusters. We run each microbenchmark N=1000 times on the same GPU and record its variation. As shown in Figure \ref{fig:var-subfigB}, we observe 0.98-6.46\% temporal variation over the microbenchmarks, highlighting the performance variability of individual kernels in a controlled environment. 

Digging further, we observe that a significant source of computation latency variation comes from the dynamically varying frequency setting of GPUs. Figure~\ref{fig:variation_sm} shows the frequency setting of the GPU streaming multiprocessor (SM) on the x-axis and the corresponding power draw of each GPU while running the microbenchmarks on the y-axis. The majority of runs see a 500 MHz difference in frequency, which corresponds to a 140W power differential. We further observe that power throttling due to thermal factors is a potential cause of observed variation. We collect NVIDIA's power throttling codes over the course of measuring N=1000 iterations of our GEMM benchmark and find that 76\% of the time, frequency changes are due to software thermal limits.

Figure \ref{fig:comm_collectives} characterizes the timing results for communication primitives. We run a series of microbenchmarks to quantify temporal and spatial variability of communication collectives, focusing on AllReduce. We find that inter-node communication collectives exhibit up to an order of magnitude higher tail latency compared to their mean, making communication a significant source of performance variation. While it is difficult to precisely measure the cause of the stragglers, particularly for small message sizes, plausible inter-node causes include network congestion and the measured compounding effects of variability and synchronization at scale. Nevertheless, software communication collectives are heavily utilized during model training, and this variability is an important factor.

\begin{figure*}[t]
      \centering    \includegraphics[width=\linewidth]{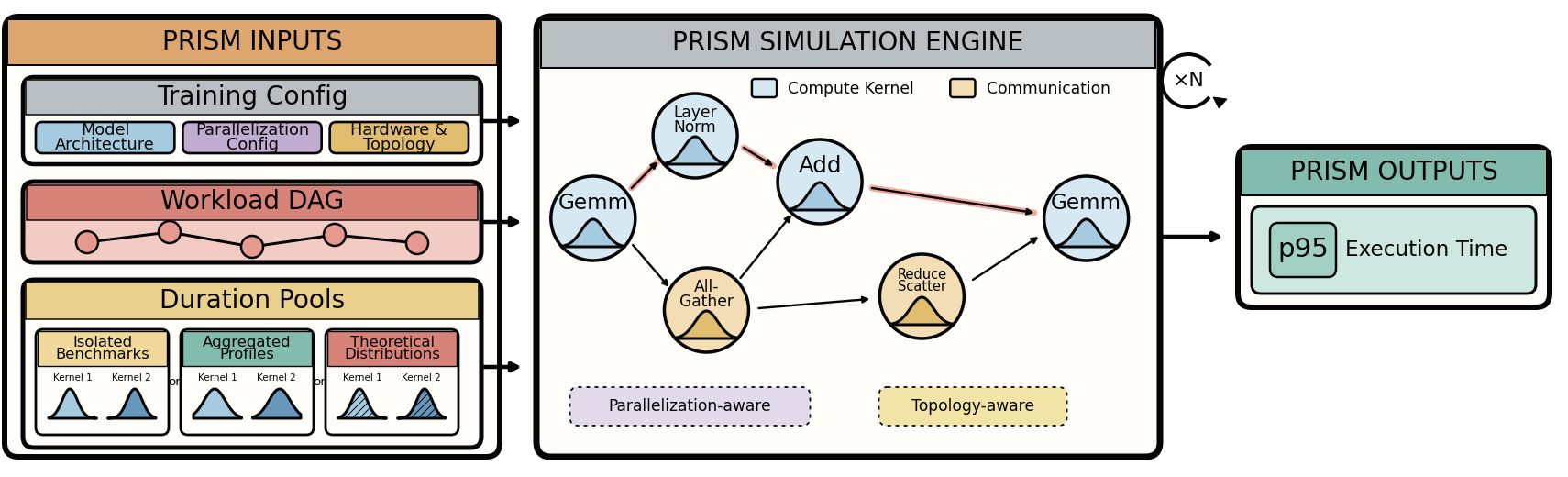}
    \caption{Overview of PRISM. Given a training configuration, multi-rank workload DAG, and operation-level duration pools, PRISM samples operation runtimes and propagates them through workload dependencies to predict p95 execution time.}
    \label{fig:PRISM_framework}
\end{figure*}

\subsection{Variability in Production Training}

To understand how runtime variability manifests in end-to-end training, we examine a production workload at 64K+ GPU scale. Figure~\ref{fig:variation_64k} characterizes variability across training steps, with p5/p50/p95 values of 17.6/19.3/21.6 seconds and a 12\% increase from the median to p95. This highlights how tail slowdowns can meaningfully increase training time even when typical step performance remains stable. Because synchronous training progresses at the rate of its slowest ranks, these slower steps directly reduce utilization and throughput. Capturing this tail behavior therefore requires modeling the runtime distribution rather than a single deterministic execution time.

These observations motivate a variability-aware performance model. Deterministic models cannot capture step-time distribution in Figure~\ref{fig:variation_64k}. PRISM addresses this gap by modeling operations as distributions to predict p95.

\section{PRISM: A Variability-Aware Performance Modeling Framework} \label{sec3}

\subsection{Design Overview} \label{sec3a}

We propose PRISM, a performance modeling framework for estimating p95 training-step time in distributed training. Unlike deterministic performance models that assign a single latency to each operation, PRISM represents compute and communication operations using runtime distributions and propagates this variability through the dependencies of a distributed training workload.

Figure~\ref{fig:PRISM_framework} gives an overview of the framework. PRISM takes three classes of input: (1) a training configuration, which describes the model, parallelization strategy, and hardware topology, (2) a multi-rank workload DAG, which captures the ordering and synchronization dependencies of execution, and (3) duration pools, which characterize the runtime variability of individual compute and communication operations. PRISM repeatedly samples from these duration pools and composes the sampled latencies through the workload DAG using Monte Carlo simulation to produce a p95 estimate.

Two properties are central to this design. First, PRISM is dependency-aware, since variability matters differently depending on where an operation lies in the execution graph and delays may be hidden by parallel work or exposed along a critical path. Second, PRISM is parallelization-aware, and the runtime distribution of an operation depends on whether it executes locally on a GPU, communicates within a node, or communicates across nodes. PRISM therefore preserves this context when constructing and sampling operation-level distributions.

\subsection{PRISM's Inputs} \label{sec3b}

PRISM represents a distributed training execution using three inputs: training configuration, workload DAG, and duration pools. We describe each in detail below. 

\textit{Training Configuration.} The training configuration provides the context required to interpret the execution graph. It specifies the model architecture, the parallelization strategy used to map the model across ranks, and the underlying hardware topology. Model architecture determines the pattern of operations in the workload. Parallelization specifies which ranks participate in each communication operation and which synchronization relationships arise during execution. State-of-the-art training typically uses a hierarchical combination of parallelisms, such as Data Parallelism (DP), Tensor Parallelism (TP), Pipeline Parallelism (PP), and Expert Parallelism (EP)~\cite{huang2019gpipe, kim2023bpipe, narayanan2021megatron, pal2019ddp, zbv, rasley2020deepspeed, yang2025cp, zhao2023fsdp, zheng2022alpa}. Hardware topology determines whether communication occurs over intra-node or inter-node links. 

\textit{Workload DAG.} PRISM represents execution as a multi-rank directed acyclic graph (DAG). Each node corresponds to a compute or communication event, while edges encode the ordering, stream, and synchronization dependencies that constrain its execution. Nodes are assigned signatures based on properties such as kernel identity, tensor dimensions, communication size, and process group. These signatures provide the link between a node in the DAG and the duration distribution from which its runtime is sampled.

\textit{Duration Pools.} A duration pool characterizes the latency distribution of a particular operation signature. PRISM treats these pools as inputs to the framework rather than quantities it must learn, and they may be obtained from (a) aggregated existing training profiles, (b) isolated operator benchmarks, or (c) fitted or user-specified statistical distributions. This design allows PRISM to model both measured systems and hypothetical operating conditions without changing the underlying simulation engine.

\begin{figure}[t]
        \centering
        \includegraphics[width=\linewidth]{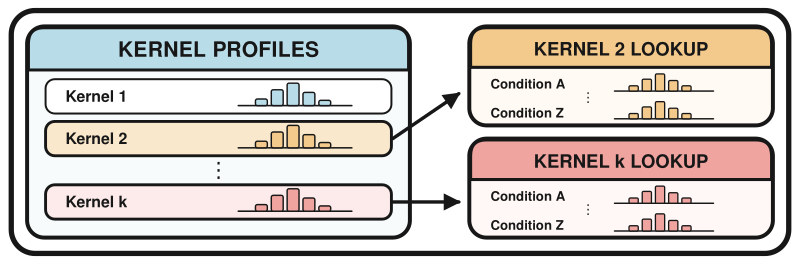}
        \caption{Duration-pool lookup in PRISM. Each operation signature can map to different runtime distributions under different execution conditions. This allows PRISM to substitute measurements from different clock frequencies, hardware platforms, or implementations while reusing the DAG.}
        \label{fig:varsubfigB}
    \label{fig:durationpool}
\end{figure}

When aggregating existing profiles, it is important to note that PRISM does not replay the measured execution. Rather, it groups observations by operation signature and generates new execution outcomes by resampling each operation during Monte Carlo simulation. Operators can also be benchmarked in isolation in order to understand how operator distributions change with hardware conditions. Figure~\ref{fig:durationpool} illustrates this abstraction. The same logical operation can be associated with different distributions under different operating conditions, such as GPU clock frequency, hardware platform, or kernel implementation. PRISM can therefore substitute one set of operation distributions for another while retaining the same workload DAG. This separation between execution structure and operation runtime is particularly useful for design-space exploration, as a user can characterize only the operations affected by a proposed system change and use PRISM to predict how that localized change propagates through the complete distributed workload.

\subsection{PRISM Simulation Engine} \label{sec3d}

Given these inputs, PRISM uses parallelization- and topology-aware methodology to produce a p95 estimate. Compute and communication are handled separately.

\paragraph{Compute.} Compute kernels such as matrix multiplications, element-wise operations, and activation functions are assigned signatures that encode properties including kernel identity and tensor dimensions. Given a profiling library for the target configuration, PRISM groups all observed occurrences with the same signature into an operation-level duration pool. For example, repeated occurrences of the same GEMM signature contribute samples to the same pool. From the duration pools that are input, grouping is done conditioned on the hardware inputs. Then, every Monte Carlo iteration draws a new duration for each DAG node from the pool selected by that node’s signature. PRISM can sample the empirical pool directly or fit it with a parametric distribution, such as normal, gamma, and lognormal distributions.

\begin{algorithm}[t]
\small
\caption{PRISM Engine}
\label{alg:prism}
\begin{algorithmic}[1]
\Require
\Statex $\Theta=(P,H,M)$
        \Comment{parallelization, hardware, model}
\Statex $G_\Theta=(V,E)$
        \Comment{causal DAG, $R$ ranks}
\Statex $\mathcal{L}_\Theta[\sigma]
         =\big\{d_{\mathrm{obs}}(v):\sigma_\Theta(v)=\sigma\big\}$
        \Comment{duration pools}
\Statex $N$
        \Comment{Monte Carlo iterations}
\Statex\hrulefill

\State $\sigma_\Theta(v)\gets\textsc{Key}\big(v,\,\Theta\big)$,
       \ \ $\forall v\in V$
       \label{ln:sig}
\State $\mathcal{L}'_\Theta[\sigma,\pi]\gets
        \textsc{Partition}\big(\mathcal{L}_\Theta[\sigma],\,H\big)$,
        \ \ $\forall$ compute $\sigma$
        \label{ln:cpupool}

\State $\mathcal{G}_\Theta\gets\textsc{Group}\big(G_\Theta,\,\Theta\big)$
       \label{ln:groups}
\ForAll{$g\in\mathcal{G}_\Theta$}
  \State $c_g\gets\tfrac{1}{|V_g|}\textstyle\sum_{i\in V_g}
          (\mathrm{end}_i-\mathrm{ready}_i)$
          \label{ln:cpool}
  \State $\mathcal{L}'_\Theta[\sigma_\Theta(g)]\mathrel{{\cup}{=}}\{c_g\}$
          \label{ln:lprime}
\EndFor

\For{$n=1$ \textbf{to} $N$}   \label{ln:loop}
  \ForAll{compute nodes $v$}
    \State $d(v)\gets\textsc{Sample}\big(d \mid \sigma_\Theta(v),\,\Theta\big)$
           \label{ln:comp}
  \EndFor
  \ForAll{$g\in\mathcal{G}_\Theta$}
    \State $c\gets\textsc{Sample}\big(d \mid \sigma_\Theta(g),\,\Theta\big)$
    \State $d(i)\gets c\ \ \forall i\in V_g$   \label{ln:shared}
  \EndFor
  \ForAll{$v\in\mathrm{topo}(G_\Theta)$}   \label{ln:sweep}
    \State $s_v\gets\tau(v)$
    \ForAll{$u\in\mathrm{pred}(v)$}
      \State $s_v\gets\max\big(s_v,\,e(u)\big)$   \label{ln:join}
    \EndFor
    \State $e(v)\gets s_v+d(v)$
  \EndFor
  \ForAll{$r=1,\dots,R$}
    \State $T^{(n)}_r\gets\max_{u\in\mathrm{pred}(\phi_r)}e(u)$
           \label{ln:readout}
  \EndFor
  \State $T^{\mathrm{step}}_{(n)}\gets\max_{r}T^{(n)}_r$
          \label{ln:steptime}
\EndFor
\State \Return $\widehat{t}_{95}
        =Q_{95}\big(\{T^{\mathrm{step}}_{(n)}\}_{n=1}^{N}\big)$
       \label{ln:p95}
\end{algorithmic}
\end{algorithm}

\paragraph{Communication.} Communication kernels including AllReduce, ReduceScatter, AllGather, AlltoAll, and Send/Recv are directly shaped by the physical network topology. Within a node, Tensor Parallel (TP) collectives communicate over high-bandwidth NVLink interconnects with low and consistent latency. Across nodes, Data Parallel (DP) and Pipeline Parallel (PP) collectives traverse the InfiniBand fabric, where they are subject to switch contention, adaptive routing, and congestion from co-located traffic. These topology-dependent effects produce duration distributions that differ meaningfully across physical placements. PRISM therefore encodes the process group (TP, DP, or CP) directly into each collective's signature, ensuring that collectives operating on different parallelism domains and traversing different physical networks are never mixed. 

A key observation underlying PRISM’s sampling design is that communication durations within a single training step are not independent: collectives and point-to-point transfers are influenced by shared network conditions, including congestion and pipeline interactions. To this end, collectives are modeled as \emph{groups}: one duration is sampled per collective invocation and broadcast to all participating ranks' DAG nodes in order to reflect the physical constraint that a collective blocks all participants equally. PRISM additionally decomposes traditional communication kernels into their \textit{arrival wait} and \textit{data transfer} phases. This is critical to understanding the impact of dependencies between compute and communication events.

\textbf{\textit{Algorithm.}} The main PRISM engine is summarized in Algorithm 1. PRISM first augments the input duration pools in order to account for hardware topology. Then, it performs Monte Carlo simulation over the input duration pools and DAG. In each iteration, PRISM samples a runtime for every operation according to the topology and correlation parameters, and propagates those durations through the DAG. Serial dependencies accumulate latency, while parallel dependencies join according to the completion time of the slowest predecessor. This process yields a distribution of execution times, which PRISM uses to calculate p95.

\subsection{PRISM's Outputs}
\label{sec:runtime_granularity}

PRISM's primary modeling target is p95 training step time. First, let \(T_{r}\) denote the completion time of rank \(r\). Then: $\mathcal{T}_s^{\mathrm{rank}}=  \left\{ T_{r, s} \right\}_{r=1}^{R}$. It follows that in synchronous training, the completion time of a step is the maximum completion time over its participating ranks, as shown:
$ T_s^{\mathrm{step}} = \max_{r} T_{s,r}.\label{eq:rank_to_step}$
PRISM can calculate p95 step time from its rank-level simulations. A complete training execution can be modeled by composing step-time samples across the training run.

\subsection{PRISM Implementation}\label{sec3e}

We implement the PRISM framework in PyTorch, including its core data structures and simulation engine components. While PRISM can operate directly on a provided DAG and corresponding duration pools along with the contextual inputs, we also build out a parsing front-end to allow for profiling of workloads, creation of duration pools, and DAG construction from a given training trace. We further integrate our framework with hooks to efficiently parse dependencies structures and kernel durations for large-scale traces. The PRISM simulation engine then leverages Monte Carlo methodology to combine the DAG and duration pools.

\section{Experimental Methodology} \label{newsec3}

Our evaluation is designed to test whether PRISM can accurately compose operation-level runtime variability through distributed execution to simulate system-level runtime distributions. We describe our methodology in this section.

\begin{table}[b]
\centering
\caption{Relative validation cost (GPU-hours) of survey distributed training performance models.}
\label{tab:validation-cost}
\footnotesize
\setlength{\tabcolsep}{4pt}
\renewcommand{\arraystretch}{1.2}

\begin{tabular}{|c|p{0.50\columnwidth}|}
\hline
\rowcolor{gray!15}
\textbf{\shortstack{\rule{0pt}{2.6ex}Validation Cost \\ (GPU-Hrs)}} & \textbf{Prior Distributed Training Works} \\
\hline \hline
$O(10)$       & \cite{calculon, habitat, daydream} \\
\hline
$O(100)$      & \cite{flexflow, amped, astrasim, galvatron, vtrain, paleo, merak, zheng2022alpa} \\
\hline
$O(1{,}000)$  & \cite{proteus, madmax, lumos, simai} \\
\hline \hline
$O(10{,}000)$ & PRISM \textit{(Ours)}, \cite{falcon} \\
\hline
\end{tabular}
\end{table}

\subsection{Dataset Construction}

We construct our evaluation dataset from 1,101 PyTorch Profiler traces collected across 14 distributed training workloads spanning a range of model architectures, parallelization strategies, and system scales. These traces serve two distinct purposes, and we separate them along independent axes. First, we reserve 344 traces for framework design and exclude them from the results reported in this paper. The remaining 757 traces support the evaluation in Section 5.

Second, and separately, within each configuration we partition the profiled training steps into an input set and a held-out set. Duration pools and the workload DAG are constructed exclusively from input steps. The measured p95 that PRISM is scored against is computed exclusively from held-out steps, which contribute nothing to PRISM's inputs. PRISM therefore never observes the step times it is asked to predict: it is given per-operation latency distributions from one set of executions and must estimate p95 of a disjoint set.

Collectively, these profiled training traces represent an \textit{order of magnitude} more GPU-hours than most prior work in distributed training modeling (see Table \ref{tab:validation-cost}). This large validation set is a key strength of the work, and distinguishes PRISM from other distributed training frameworks. This investment in experimental scale ensures that PRISM's tail estimates are accurate across a wide range of model architectures and configurations.

\subsection{Evaluation Metrics} \label{sec:validation_protocol}

We evaluate PRISM along two dimensions: \emph{coverage} and \emph{scale}. Coverage measures whether PRISM accurately captures runtime distributions across different model architectures, parallelization strategies, and execution structures. Scale measures whether this accuracy is maintained as distributed training grows to larger GPU counts. Section~5.3 separately evaluates PRISM when its duration distributions are obtained from alternative sources, including isolated operator measurements.

Our primary quantitative target is the 95th-percentile runtime. For each configuration, we compare PRISM's estimated p95, against the p95 measured from the held-out executions. We report relative p95 error as:
$
    \mathrm{P95\ Error}\;
    =
    100
    \left(
    \frac{Q_{95}^{\mathrm{PRISM}}}
         {Q_{95}^{\mathrm{Measured}}}
    - 1
    \right).
    \label{eq:p95_error}
$

\begin{figure}[t]
    \centering
        \centering
        \includegraphics[width=0.99\linewidth]{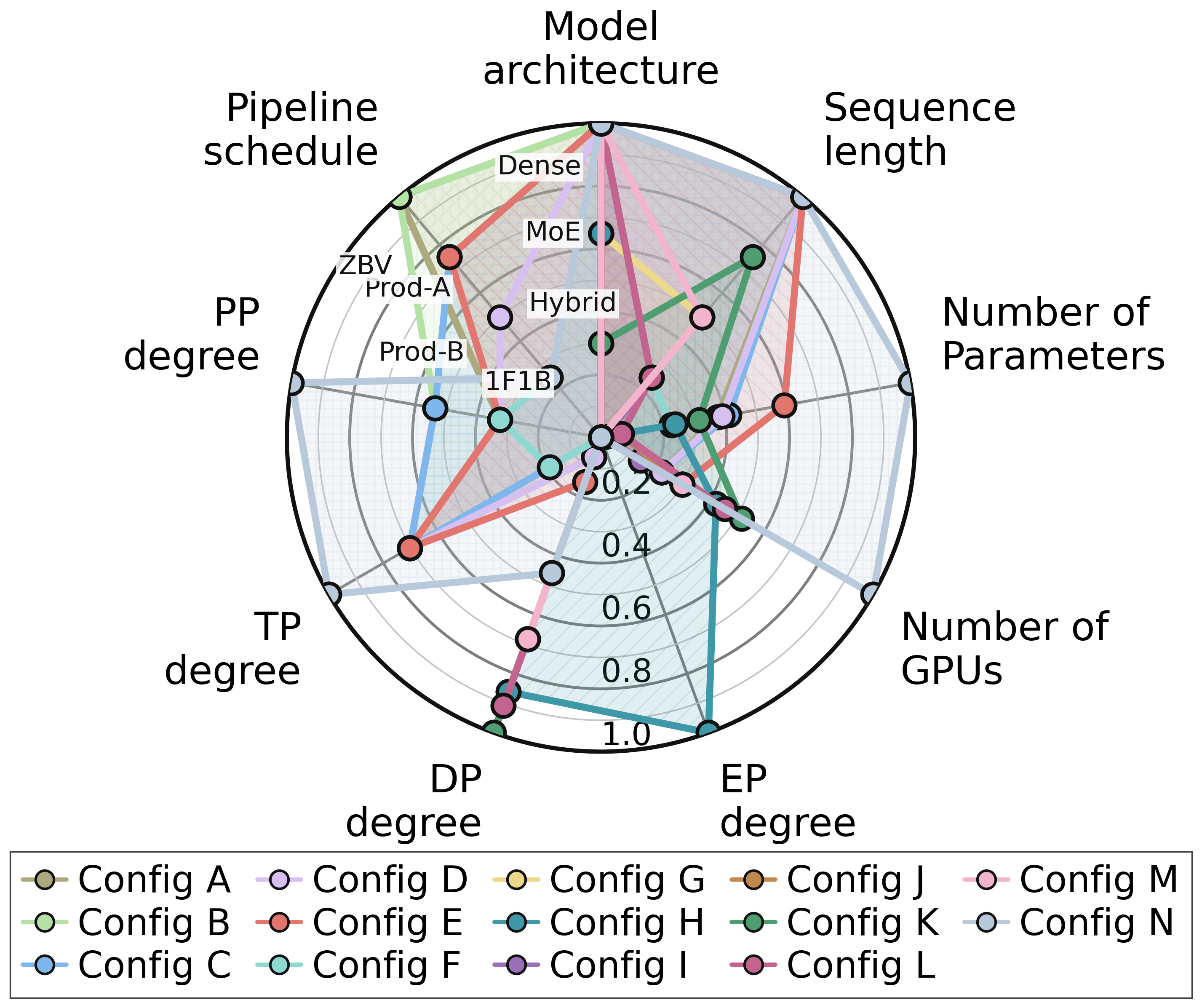}
        \caption{PRISM Evaluation Suite. We evaluate PRISM across 14 real-world training configurations spanning dense Transformer, MoE, and hybrid/SSM architectures; model sizes from 0.73B–2.2T parameters; sequence lengths from 2K–32K; multiple DP, TP, and PP configurations and pipeline schedules; and deployments ranging from hundreds to 64K+ GPUs.}
        \label{fig:varsubfigC}
\end{figure}

\subsection{Simulation Configuration}

Unless otherwise stated, PRISM uses $N=1{,}000$ Monte Carlo iterations per configuration. In each iteration, PRISM samples operation durations using the topology- and parallelization-aware procedure described in Section~3.3 and propagates those samples through the workload DAG to obtain an execution outcome. We empirically verify that the estimated p95 converges by $N=1{,}000$. Increasing the number of Monte Carlo iterations beyond this point produces negligible changes in the reported estimates.

\section{Evaluation Results for PRISM} \label{sec4}

\subsection{PRISM Evaluation Suite} \label{sec4a}

We now evaluate PRISM across a suite of 14 real-world distributed training workloads. As summarized in Figure \ref{fig:varsubfigC}, the evaluation suite is assembled to vary properties that affect PRISM's estimation: (i) model architecture diversity, (ii) parallelization configurations, and (iii) physical scale.

\textit{Model Architecture Diversity}. The evaluation suite contains models ranging from 0.73B to 2.2T parameters and includes dense Transformer, mixture-of-experts (MoE), and hybrid/SSM architectures. These models expose different computational structures, including dense and sparse Transformer blocks, feed-forward layers, and SSM operators. We additionally evaluate sequence lengths from 2,048 to 32,768 tokens, which impacts both compute kernel dimensions and communication kernel message sizes.

\textit{Parallelization Configurations}. The workload suite additionally contains combinations of various parallelization configuration across data parallelism (DP), tensor parallelism (TP), expert parallelism (EP), and pipeline parallelism (PP), which changes both the shape of the execution DAG and the communication primitives exercised by the workload. We include training configurations with different pipeline schedules as well, including ZBV and two production pipeline schedules (Prod-A and Prod-B). These choices alter the workload DAG, communication groups, synchronization points, and critical-path structure that PRISM must model.

\textit{Physical Scale}. Finally, our evaluation suite contains jobs ranging from controlled configurations operating on hundreds of GPUs to production training workloads spanning up to 64K+ GPUs. This diversity stresses hardware participating in execution, as well as the opportunities for rank imbalance, communication contention, and stragglers.

\begin{figure}[t]
        \centering
          \includegraphics[width=\linewidth]{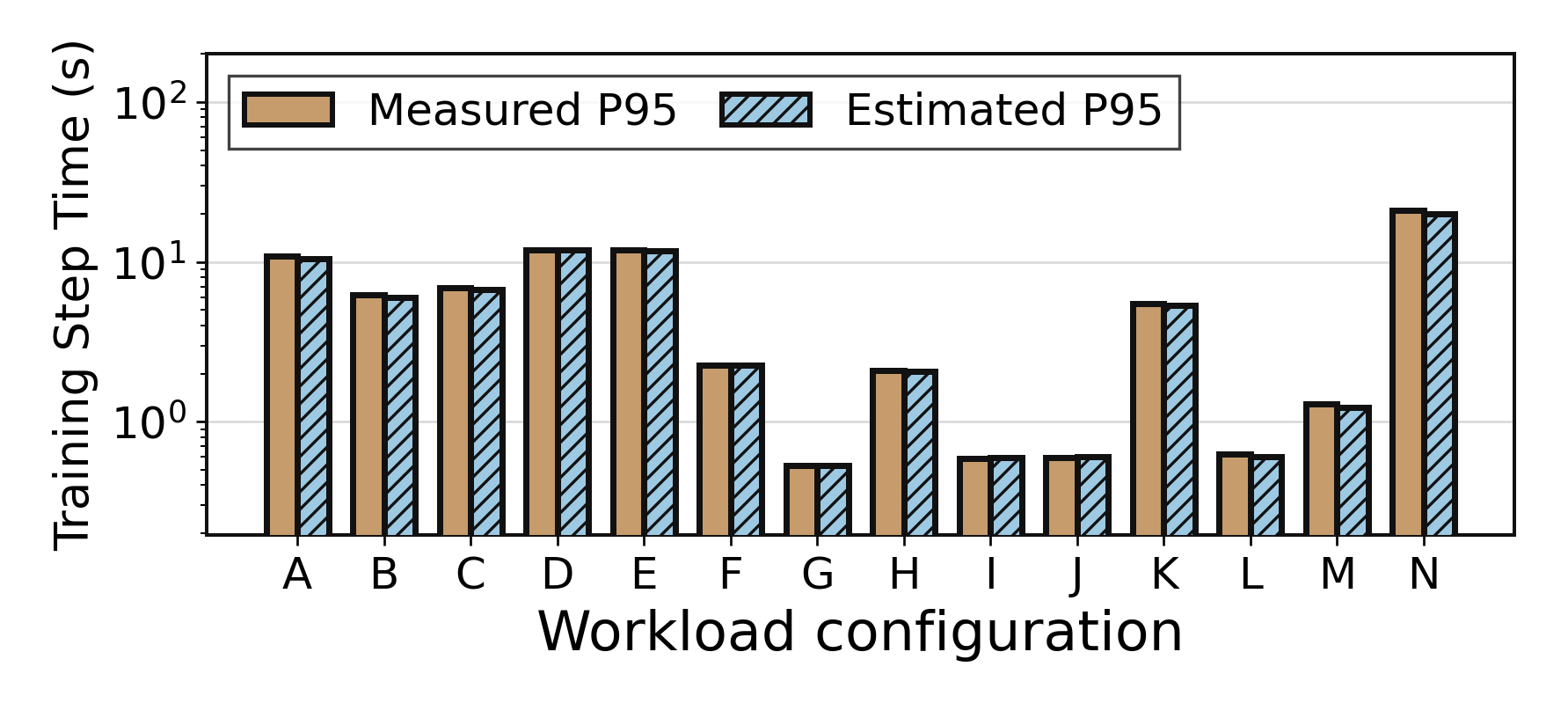}
        \caption{PRISM accurately estimates P95 training step time within 5.4\% error across the diverse set of 14 real-world workload configurations, as shown in Figure \ref{fig:varsubfigC}.  }
        \label{fig:validation1}
    \label{fig:validation}
\end{figure}

\subsection{Quantifying Tail-Estimation Accuracy}

\begin{figure}[t]
        \centering
        \includegraphics[width=\linewidth]{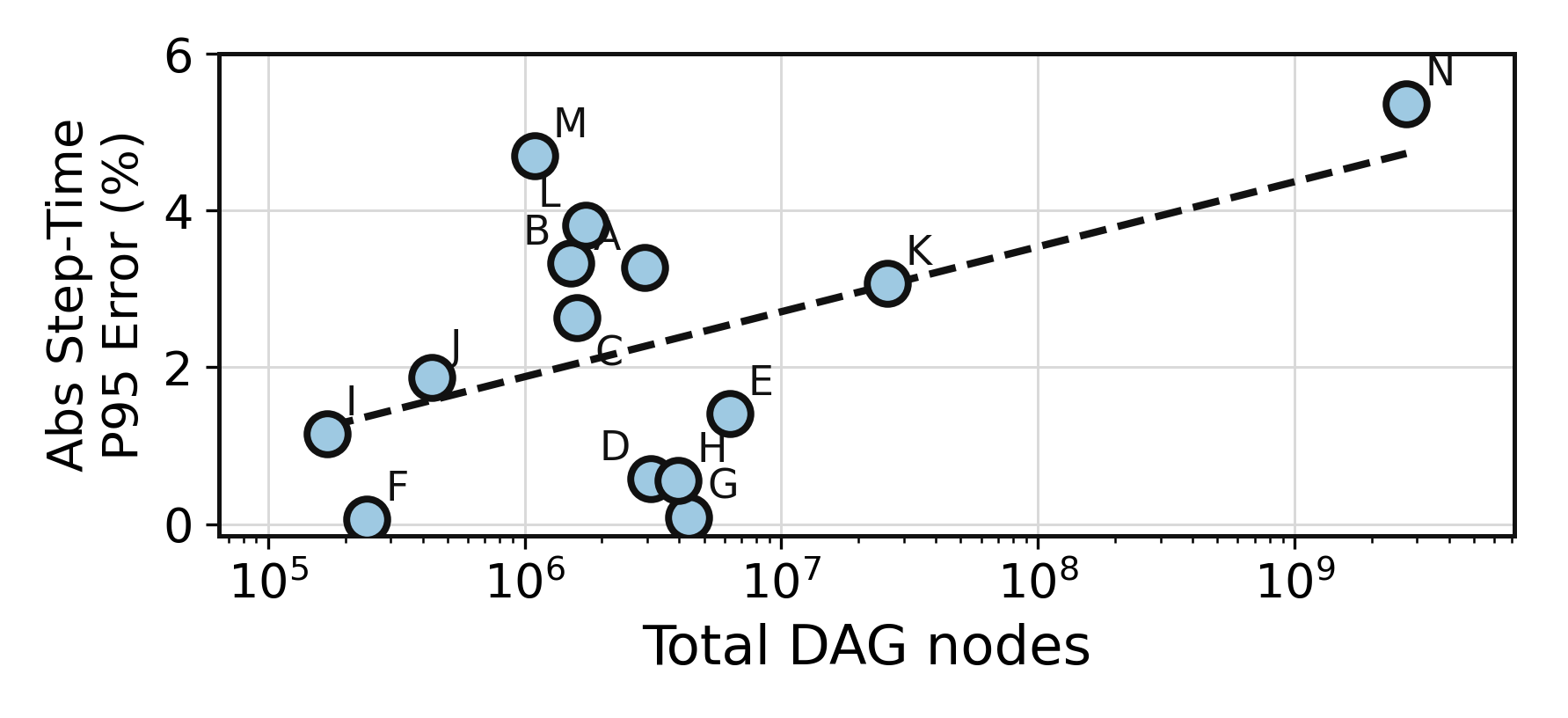}
        \caption{PRISM remains accurate as workload DAGs grow in size. Absolute P95 error increases slightly with DAG node count (Spearman \(\rho=0.14\)) but remains below 5.4\% across workloads spanning orders of magnitude in graph size.}
        \label{fig:varsubfigB}
    \label{fig:validation2}
\end{figure}

Figure~\ref{fig:validation1} quantifies PRISM's tail estimation accuracy across the evaluation suite. We evaluate PRISM on its primary modeling target: p95 training step time. As described in Section~\ref{newsec3}, we compare PRISM's estimated p95 execution time against the p95 measured from real training runs. Across the full suite (x-axis), PRISM estimates p95 execution time across held-out steps, with a maximum of 5.4\% error and 2.3\% mean error (y-axis), demonstrating that operation-level latency distributions can be composed through a distributed training DAG to accurately recover end-to-end tail performance.

To examine how PRISM's accuracy evolves across the breadth and scale of the workload suite, we examine how estimation error changes as the workload execution grows in complexity. As DAGs grow, PRISM must compose variability across a larger number of compute operations, communication events, synchronization points, and competing execution paths. Figure~\ref{fig:validation2} plots absolute p95 estimation error on the y-axis against the total number of nodes in each workload DAG on the x-axis. Across the 14 configurations, we observe a slight positive relationship between DAG size and estimation error (Spearman $\rho$ = 0.14), We observe all configurations remain within 5.4\% error even as DAG size spans several orders of magnitude. Thus, although errors accumulate moderately with graph complexity, PRISM remains accurate for large distributed execution graphs.

\subsection{Composing Isolated Measurements at Scale}

An additional advantage of PRISM is that it avoids requiring repeated execution of a complete distributed workload when a user wants to evaluate a new system configuration. Instead, PRISM can characterize individual operations and compose those measurements through the workload DAG to estimate the behavior of the full distributed execution. 

\begin{figure}[t]
    \centering
        \centering
         \includegraphics[width=0.99\linewidth]{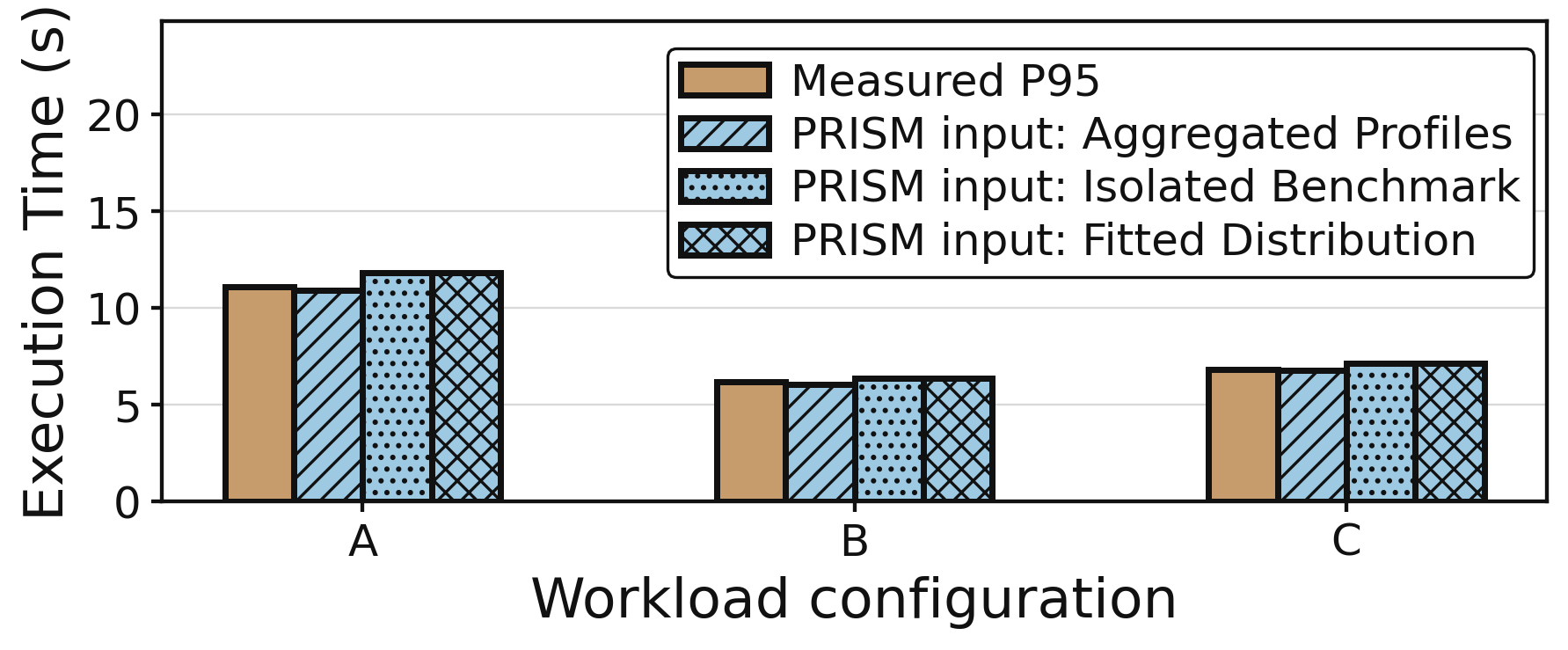} 
        \caption{PRISM's predicted p95 training step time under three different duration-pool sources referenced in Section~\ref{sec3b}. Despite replacing end-to-end kernel distribution profiles with isolated operator measurements, PRISM estimates p95 within 6.36\%.}
    \label{fig:validation3b}
\end{figure}

Figure~\ref{fig:validation3b} evaluates how various sources of measurements can be used to estimate p95. We compare PRISM estimations using three sources of duration pools: durations extracted from workload profiles, isolated benchmarks, and fitted distributions. In all three cases, the resulting duration pools are substituted into the same workload DAG and PRISM composes them with the workload's structure to estimate the resulting p95 step time with a maximum error of 6.36\%. 

This property is particularly important for design-space exploration. Consider evaluating a new kernel implementation, GPU operating point, or hardware configuration. Rather than executing multiple training runs under every candidate design, a user can benchmark the affected operations, replace their duration pools in PRISM, and retain the existing workload structure and unaffected operation distributions. PRISM then estimates how those localized changes propagate through the large distributed execution. Section 6.1 additionally validates PRISM against a measured hardware configuration it was not profiled on.
\section{Leveraging PRISM for Variability-Aware System Design} \label{sec5}

Having established that PRISM estimates p95 execution time within 5.4\% error, we now use the model to derive insights into efficient system-design. In particular, we leverage PRISM-Guide, which turns PRISM's tail performance predictions into variability-aware system-design recommendations and insights. PRISM-Guide explores candidate scenarios using 
PRISM and ranks them according to their estimated p95 performance.

\begin{figure}[t]
        \centering
        \includegraphics[width=\linewidth]{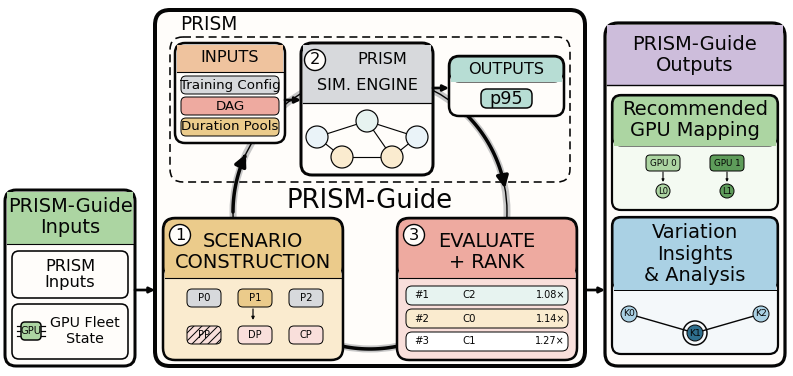}
        \caption{Overview of PRISM-Guide. PRISM-Guide turns PRISM's tail performance estimation into variability-aware system-design recommendations. PRISM-Guide combines PRISM's workload
inputs with the current hardware state, constructs candidate system scenarios,
and evaluates each candidate using PRISM. Candidates are ranked by their p95 estimate to produce configuration recommendations or identify insights regarding runtime variability.}
    \label{fig:prismguide}
\end{figure}

Figure~\ref{fig:prismguide} illustrates this workflow. PRISM-Guide combines PRISM's workload inputs with a description of   the current hardware or fleet state. This consists of the telemetry of each GPU that is marked for a certain workload execution, and can be collected via a simple profiling script. It constructs candidate scenarios corresponding to feasible system 
decisions, evaluates each scenario with PRISM, and ranks the resulting p95 
estimates. The output can be used either to recommend a system configuration 
or to identify variability insights. We demonstrate both uses below.

\subsection{Recommending Placement of a Degraded GPU} \label{sec5a}

\begin{figure*}[t]
    \centering
    \includegraphics[width=\linewidth]{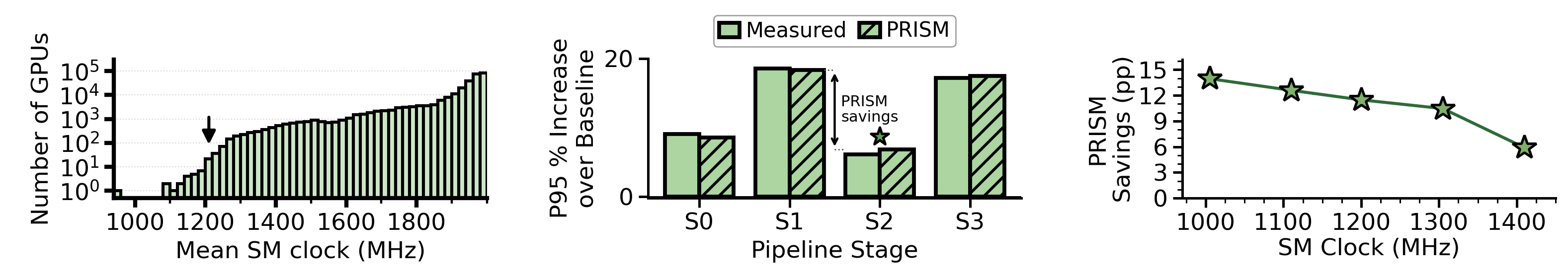}
    \label{fig:ablation_b}
    \caption{PRISM-Guide recommends where to place a degraded GPU.
    \textit{Left:} Distribution of mean H100 SM frequency across 286,212 GPUs. In this example, we model one GPU at 1200\,MHz while the remaining nodes operate nominally.  \textit{Middle:} Measured and PRISM-estimated p95 increase over
the nominal baseline when the degraded GPU is placed in each pipeline stage. PRISM-Guide recommends S2, avoiding up to an 11.47 percentage point p95 increase relative to worst placement. Measured results confirm same best placement. \textit{Right:} Maximum p95 savings from
PRISM-Guide's placement recommendation as the degraded GPU clock is swept from
1005 to 1410 MHz. Placement savings range from 5.89 to 13.94 percentage points.}
    \label{fig:prismguide1}
\end{figure*}

We first use PRISM to quantify how GPU performance degradation propagates through distributed execution. We model a common production scenario in which GPUs enter a lower-performance operating state because of power, thermal, or other datacenter conditions. We ground the degraded-node scenario using measurements from our fleet. Figure~\ref{fig:prismguide1} (left) shows the distribution of mean SM frequency across 286,212 GPUs operating at 100\% utilization. Observed operating 
points span approximately 945--1980 MHz. We select 1200 MHz as a representative degraded operating point, shown by the arrow, and model one GPU at this frequency while the remaining GPUs operate nominally.

Given this hardware state, PRISM-Guide constructs one candidate configuration for each possible pipeline-stage placement of the degraded GPU. The same 1200 MHz GPU is assigned to pipeline stages S0 through S3 in turn, while the workload, remaining nodes, and all other configuration parameters are held fixed. PRISM predicts the p95 execution-time penalty for each candidate and ranks the placements.

Figure~\ref{fig:prismguide1} shows the result, with pipeline stages on the x-axis and p95 percent increase on the y-axis. Including the degraded GPU increases p95 by 6.87--18.34\%, depending on the pipeline stage it is placed in. As indicated by the star, PRISM-Guide recommends placing the degraded node in stage S2, which has the lowest estimated p95 increase among the candidate mappings. Leveraging PRISM-Guide's recommendation saves up to 11.47 percentage point p95 increase as compared to worst placement. We compare this with measured results to validate the recommendation. We leverage sudo access to degrade a set of GPUs to 1200 MHz and force a logical-physical GPU mapping to execute each placement experimentally. We validate the measured results preserve the same ranking as PRISM estimates, confirming S2 is the best observed placement for this configuration.

The benefit of placement also depends on the severity of the degradation. 
Figure~\ref{fig:prismguide1} (right) sweeps the degraded GPU from 
1005 to 1410 MHz for the same workload. As the GPU becomes slower, the 
performance difference between candidate placements grows, and the maximum savings 
from selecting the PRISM-Guide placement ranges from 5.89 to 13.94 percentage 
points. Thus, placement becomes increasingly important as hardware heterogeneity 
increases.

\begin{figure}[b]
    \centering
    \includegraphics[width=0.9\linewidth]{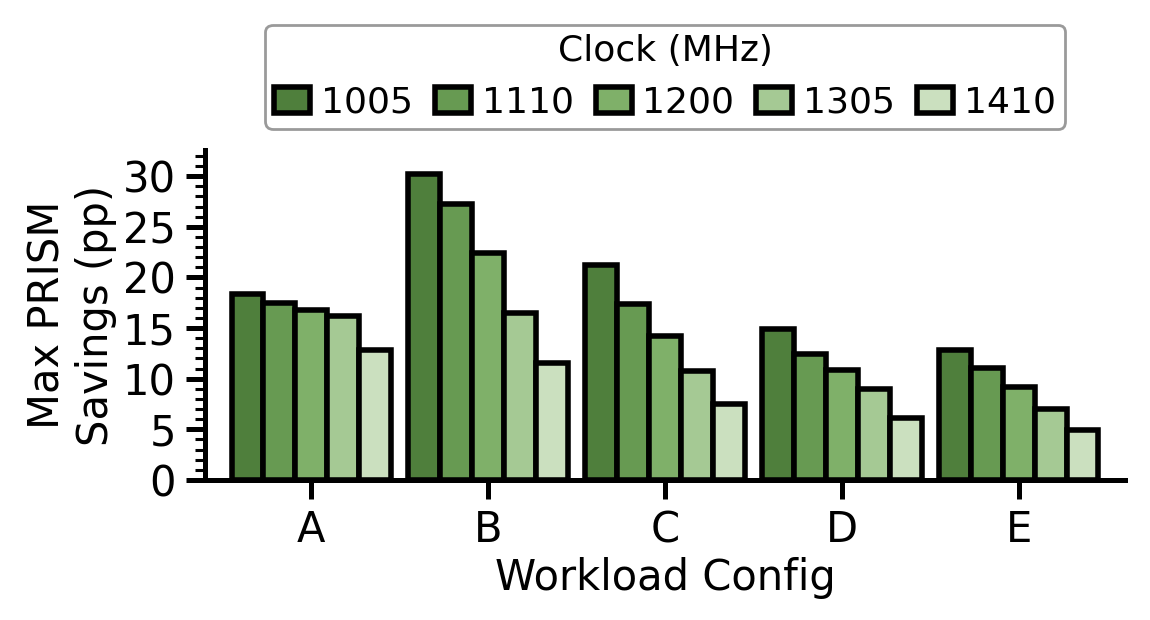}
    \label{fig:ablation_b}
    \caption{PRISM-Guide placement recommendations across workloads and
degradation levels. Maximum p95 savings from selecting the best slow-GPU
placement across five workload configurations and five degraded-GPU
clock frequencies. Savings increase as the GPU slows and vary
with workload, ranging from 4.9 to 30.2 percentage points.}
    \label{fig:prismguide2}
\end{figure}

We additionally sweep across different workload configurations. Figure~\ref{fig:prismguide2} repeats the placement analysis across five 
workloads and five degraded-GPU clock frequencies. The benefit is workload-dependent, but the same overall trend holds, and lower clock frequencies increase the importance of placement. Across the sweep, PRISM-Guide identifies 
placement opportunities worth 4.9--30.2 percentage points of p95 improvement. 

These results show that degraded GPUs need not always be removed from a training allocation. By accounting for how a degraded GPU interacts with the workload’s execution structure, PRISM-Guide can identify placements that substantially reduce its tail-performance impact, enabling systems to continue using otherwise functional GPUs operating in degraded states.

\subsection{Providing Variation Insights and Root Cause Analysis} \label{sec5b}

\begin{figure}[b]
    \centering
     \begin{subfigure}{0.49\linewidth}
        \centering
        \includegraphics[width=\linewidth]{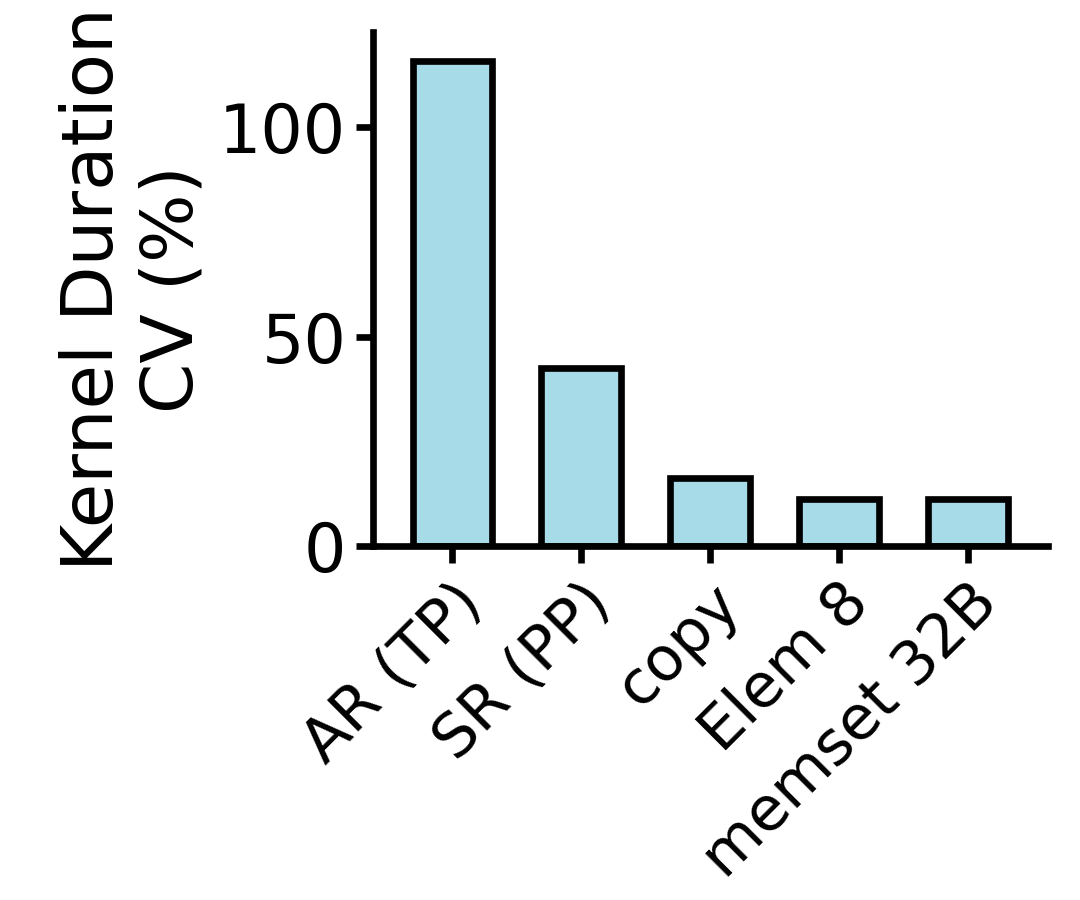}
        \label{fig:ablation_c}
    \end{subfigure}
    \medskip 
    \begin{subfigure}{0.49\linewidth}
        \centering
        \includegraphics[width=\linewidth]{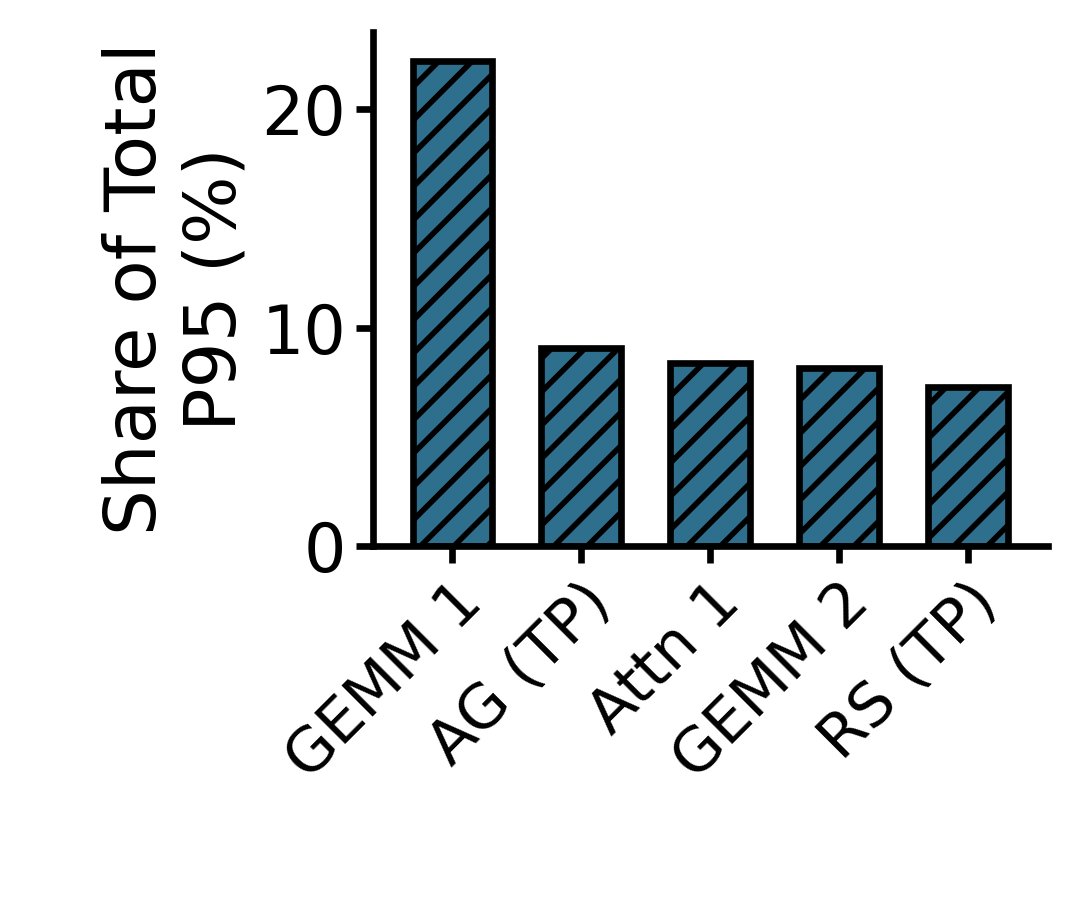}
        \label{fig:ablation_b}
    \end{subfigure}

    \caption{PRISM-Guide distinguishes local kernel variability from
end-to-end performance impact. \textit{Left:} Operations with the largest
kernel-duration coefficient of variation (CV). \textit{Right:} The top five
kernels with impact on p95 execution time identified through PRISM. We find high kernel variability does not necessarily imply high system impact, motivating PRISM's dependency-aware insights.}

    \label{fig:prismguide3}
\end{figure}

PRISM-Guide can also leverage PRISM to provide insights on top contributors to variation. A key advantage of PRISM is that it does not treat p95 step time as a black-box. PRISM leverages kernel-level execution models to isolate where variability originates and models its propagation through a given workload. This is important since a kernel’s impact depends jointly on its variability, duration, and position within the DAG, and can provide insights into where variability-reduction effort is likely to have the greatest impact.

To quantify this downstream impact, PRISM-Guide performs controlled source
ablations. For each candidate operation, we preserve that operation's measured
runtime distribution while holding all other operation durations fixed. Figure~\ref{fig:prismguide3} compares operation-level variability with
its downstream effect. The left panel ranks kernels by coefficient of variation
in their local runtime distributions, while the right panel ranks operations by
their impact on end-to-end p95 under PRISM-Guide's source ablation.
The rankings differ substantially. For example, GEMM1 and tensor-parallel
communication operations appear among the largest contributors to p95 even
though the operations with the highest local CV are different.

This difference demonstrates why kernel-level variance alone is insufficient
for prioritizing optimization. We
further validate this behavior experimentally by removing the source of an
NVLink degradation and observe AllGather falls from the
second-largest contributor to the fifth, consistent with the reduction in
communication variability. PRISM-Guide accounts for how that variance is exposed through the workload's synchronization and dependency structure.
\section{Related Work}

\subsection{Large-Scale Training and Performance Modeling}
Performance modeling for large-scale training has traditionally relied on analytical abstractions. 
Systems such as~\cite{zhang2024megascale, metis, amped, astrasim, merak, zheng2022alpa} and related works focus on optimizing parallelism strategies under ideal conditions. For example,~\cite{madmax} introduces a trace-based analytical model to project optimal parallelization strategies while assuming consistent hardware performance regardless of the underlying workload.
Although effective for making high-level design decisions, these models assume uniform hardware behavior and fail to capture the real-world variability that emerges at scale.

Other existing frameworks, such as~\cite{lumos, vtrain} offer detailed profiling capabilities but focus primarily on average case performance.
This limits the framework's ability to model training dynamics in real-world clusters, where mixed GPU generations, thermal variation, and non-uniform interconnects are increasingly common.
In contrast, PRISM explicitly models these variations in both space and time, allowing more accurate and realistic predictions for modern and future training workloads. 

Deterministic performance models and PRISM therefore address fundamentally different targets. Existing simulators estimate a single nominal or expected runtime from fixed operation costs, whereas PRISM estimates p95 step time under empirically observed runtime variability. A direct numerical comparison would require treating a deterministic point estimate as a tail prediction, despite that estimate being neither designed nor calibrated for p95 behavior. PRISM is therefore complementary to these frameworks, extending performance modeling to capture how operation-level latency distributions, correlated communication behavior, and workload dependencies compose into tail performance. 

\subsection{Training Disruptions}
A separate body of work has investigated failures in training clusters at-scale, characterizing their frequency of occurrence in production clusters.
Papers such as~\cite{kokolis2025reliability} break down the causes of failures and assign probabilities for various root causes. Some recent works have quantified variation~\cite{notAllGpuscreatedequal} and fail-slows at-scale as well~\cite{lin2025stragglers}.
Most works in this area, such as~\cite{lin2025stragglers, falcon, holmes}, propose \textit{monitoring} tools to keep track of slowdowns so that manual intervention can occur. Others also present \textit{mitigation} strategies~\cite{falcon}. This is orthogonal to our work, which aims to develop a \textit{modeling} and \textit{design-space exploration} tool for modeling variability at-scale.  Given that practitioners are often not able to pinpoint where variability originates, its important to model and build systems that perform well despite this observed variation. PRISM is designed to help researchers understand and explore the design space of variation's impact at-scale.

\section{Conclusion}

Through this work, we introduce PRISM, a variability-aware framework for modeling large-scale training. We motivate p95 as a practical tail-performance target and demonstrate that PRISM estimates within 5.4\% error across a suite of 14 diverse training configurations. More broadly, PRISM re-frames variability as an important system-design signal through PRISM-Guide. This provides a principled foundation for placement decisions and optimizations that help deliver robust, predictable performance at scale. As distributed training continues to scale, explicitly accounting for runtime variability will become increasingly important for both performance modeling and system design.

\section{Acknowledgments}

We write the text of the paper ourselves, but use AI to edit and help rephrase, condense, and clarify throughout.

\bibliographystyle{ACM-Reference-Format}
\bibliography{software}


\end{document}